\pdfoutput=1
\documentclass[table,nofootinbib,secnumarabic]{revtex4-2}
\usepackage{epsf,amsmath,amssymb,graphicx,dcolumn}
\usepackage{scalefnt}
\usepackage{hyperref}
\usepackage{cleveref,etoolbox,color,soul}
\usepackage{listings,booktabs}
\usepackage[utf8]{inputenc}
\usepackage{epsfig}
\usepackage{graphicx}
\usepackage[colorinlistoftodos]{todonotes}
\usepackage{multirow}

\usepackage{mathtools}

\usepackage{slashed}

\definecolor{linkcolor}{rgb}{0,0,0.5}

\def\be{\begin{equation}}
\def\ee{\end{equation}}
\def\ba{\begin{align}}
\def\ea{\end{align}}
\def\bea{\begin{eqnarray}}
\def\eea{\end{eqnarray}}

\def\L{{\rm L}}
\def\R{{\rm R}}
\def\GeV{{\rm GeV}}

\begin{document}

\title{A closer look at the $U(1)_{B-L}$ explanation of the ATOMKI nuclear anomalies}

\author{P.M.~Ferreira}
\email{pmmferreira@fc.ul.pt}
\affiliation{Instituto Superior de Engenharia de Lisboa, Instituto Polit\'ecnico de Lisboa 1959-007 Lisboa, Portugal}
\affiliation{Centro de F\'{\i}sica Te\'orica e Computacional, Faculdade de Ci\^encias,
Universidade de Lisboa, Campo Grande, Edif\'{\i}cio C8 1749-016 Lisboa, Portugal}
\author{B.L.~Gon\c{c}alves}
\email{bernardo.lopes.goncalves@tecnico.ulisboa.pt}
\affiliation{Departamento de F\'{\i}sica and CFTP, Instituto Superior T\'ecnico, Universidade de Lisboa, Lisboa, Portugal}
\affiliation{Centro de F\'{\i}sica Te\'orica e Computacional, Faculdade de Ci\^encias,
Universidade de Lisboa, Campo Grande, Edif\'{\i}cio C8 1749-016 Lisboa, Portugal}
\author{F.R.~Joaquim}
\email{filipe.joaquim@tecnico.ulisboa.pt}
\affiliation{Departamento de F\'{\i}sica and CFTP, Instituto Superior T\'ecnico, Universidade de Lisboa, Lisboa, Portugal}

\begin{abstract}
We revisit the gauged $U(1)_{B-L}$ explanation of the ATOMKI nuclear anomalies, in which the new gauge boson is the hypothetical $X(17)$ particle. It is known that the vanilla $B-L$ scenario is unable to account for appropriate couplings, namely the suppression of the couplings of $X(17)$ to neutrinos, which motivates adding vector-like leptons. The simplest case, in which the new fields have $B-L$ charges equal to $1$, is highly disfavoured since it requires large mixing with the Standard Model fields. One solution recently put forward is to consider large $B-L$ charges to counterbalance small mixing. We show that, in this scenario, and after taking into account several phenomenological constraints, the dominant contribution to the muon anomalous magnetic moment $(g-2)_\mu$ is expected to be extremely large and with a negative sign, being thus excluded by experiment.
\end{abstract}

\maketitle

\section{Introduction}

The ATOMKI pair spectrometer experiment was set up to look for $e^{+}e^{-}$ internal pair creation (IPC) in the decay of excited ${{}^{8}\text{Be}}$ nuclei. When collecting data, a possible New Physics (NP) hint was observed in the decay of the excited state with spin-parity $J^P=1^{+}$ into the ground state $0^{+}$, both with isospin $T=0$, and with an excitation energy of $18.15~\text{MeV}$ -- a significant enhancement of IPC was observed at large angles in the angular correlation of this transition~\cite{Krasznahorkay:2015iga}. It was subsequently pointed out that the observed excess was consistent with an intermediate light boson $X$ (produced on-shell in the decay of the excited state) which then decays into a $e^{+} e^{-}$ pair. Improved investigations constrained the mass of the hypothetical new particle to be $m_X \simeq 17~\text{MeV}$. More recently, new anomalies reported by the same collaboration were observed in ${{}^{4}\text{He}}$~\cite{Krasznahorkay:2021joi} and ${{}^{12}\text{C}}$~\cite{Krasznahorkay:2022pxs}, which support the existence of the hypothesised $X(17)$ vector boson~\cite{Feng:2020mbt} 
-- for a recent summary of the status on the $X(17)$ boson, see~\cite{Alves:2023ree}.

Explanations for the nature of $X(17)$ were put forward in Refs.~\cite{Feng:2016jff} and \cite{Feng:2016ysn}, considering a new vector boson $Z^\prime$ with a mass of about $17~\text{MeV}$. In the former work, the ATOMKI results were combined with other experimental data to constrain, in a model-independent way, the couplings of the $X$ boson to fermions. One of the main conclusions extracted from this analysis is that, based on searches for $\pi^0 \rightarrow Z^\prime + \gamma$ by the NA48/2 experiment~\cite{NA482:2015wmo}, the $X$ particle should couple more to neutrons than to protons, so that the $X$ particle was dubbed to be ``protophobic". In~\cite{Feng:2016ysn}, simple $U(1)$ extensions of the Standard Model (SM) were discussed that can account for a $X(17)$ having the required couplings to explain the anomaly\footnote{A plethora of models has appeared in the literature -- as an example, see~\cite{DelleRose:2017xil}, in which the ATOMKI anomaly is addressed in an extension of the SM with a gauged $U^\prime(1)$ symmetry in the context of the two Higgs doublet model (2HDM).}. One of the proposed scenarios relies on a $U(1)_{B-L}$ extension of the SM gauge group which, however, suffers from the fact that the $X$ couplings to neutrinos turn out to be too large, as shown in Ref.~\cite{Feng:2016ysn}. According to the authors of that reference, this problem can be circumvented by adding vector-like leptons (VLLs) that mix with SM leptons. This framework was later considered in~\cite{Hati:2020fzp} to provide a combined explanation of both the ATOMKI and the $(g-2)_{e,\mu}$ anomalies. A similar goal was pursued, for instance, in~\cite{Nomura:2020kcw} within a $U(1)_X$ extended flavour violating 2HDM. 

The mixing between SM leptons and the new degrees of freedom stemming from the presence of VLL fields can be \textit{a priori} completely general. Still, severe phenomenological constraints may apply. For instance, in Refs.~\cite{Chun:2020uzw,Dermisek:2021ajd}, the $(g-2)_{e,\mu}$ anomalies were considered within the 2HDM extended with VLLs that mix with the SM. Since the mixing between SM and vector-like fields occurs through the coupling with a $SU(2)$ scalar doublet, $Z$-pole measurements require small mixing. On the contrary, and in order to justify the suppression of neutrino-$X$ couplings, the mixing considered in~\cite{Feng:2016ysn} must be sizeable. This is possible to achieve since SM fields only mix with new leptons that carry the same SM quantum numbers, due to the presence of an extra scalar singlet. In this case, that mixing is not constrained by precision measurements. However, one would expect mixing with the SM to be constrained by other measurements, such as those on Higgs coupling modifiers or neutrino unitarity constraints, for which significant deviations have not yet been found.

Recently, an up-to-date analysis was conducted in~\cite{Denton:2023gat}, taking into account the anomalies seen in ${{}^{8}\text{Be}}$, ${{}^{4}\text{He}}$ and ${{}^{12}\text{C}}$, simultaneously. The authors found out that, while the different measurements do  not perfectly agree with each other, the tension mostly depends on uncertain isospin effects, which are less significant than the overall hint for NP. It was also found that, in order to comply with neutrino unitarity constraints, large mixing between the SM and non-SM leptons is disfavoured,  and that the suppression of neutrino couplings to $X$ is possible through large $B-L$ charges.

In this paper, we revisit the ATOMKI nuclear anomalies in the context of the aforementioned $B-L$ model with VLLs, summarised in Section II. In Section III we briefly recap the results from~\cite{Feng:2016ysn} and~\cite{Denton:2023gat}, extending their discussion by looking at the second lepton generation, and at the Higgs sector in more detail. In particular, we set limits on the mass of the new scalar particle present in the model, and consider the present constraints on SM Higgs couplings to muons set by the CMS experiment~\cite{CMS:2022dwd}. We check whether the neutrino couplings can be consistently suppressed given all those constraints. Taking all aspects into account, we reevaluate the NP contributions to $(g-2)_\mu$ to show that, in the scenario where $X(17)$ still explains the ATOMKI anomaly, the dominant contribution to $(g-2)_\mu$ is solidly ruled out by experiment~\cite{Muong-2:2023cdq}. We draw our final conclusions in Section IV.

\section{Recapping the $B-L$ model with VLLs} 
\begin{table}[]
    \centering
   \begin{tabular}{|c c|| c c c c c|} 
 \cline{2-6}
  \multicolumn{1}{c|}{} & \rule{0pt}{3.0ex} Fields \, & \, $SU(3)_c$ \, & \, $SU(2)_{\rm L}$ \, & \, $U(1)_Y$ \, & \, $Q_{B-L}$ & \multicolumn{1}{|c}{} \\ [0.5ex] 
\hline  
  \multicolumn{1}{|c|}{} & \rule{0pt}{3.0ex} ${q_i}_\L = ({u_i}_\L, {d_i}_\L)^T$ & $\mathbf{3}$ & $\mathbf{2}$ & $1/6$ & $1/3$ & \multicolumn{1}{|c|}{} \\ [0.5ex] 
 \cline{2-6}
 \multicolumn{1}{|c|}{\parbox[t]{5.5mm}{\multirow{3}{*}{\rotatebox[origin=c]{90}{Vanilla $B-L$ \hspace{5.0mm}}}}} & \rule{0pt}{3.0ex} ${u_i}_\R$ & $\mathbf{3}$ & $\mathbf{1}$ & $2/3$ & $1/3$ & \multicolumn{1}{|c|}{} \\ [0.5ex] 
 \cline{2-6} 
  \multicolumn{1}{|c|}{} & \rule{0pt}{3.0ex} ${d_i}_\R$ & $\mathbf{3}$ & $\mathbf{1}$ & $-1/3$ & $1/3$ & \multicolumn{1}{|c|}{\parbox[t]{5.5mm}{\multirow{3}{*}{\rotatebox[origin=c]{90}{Vanilla $B-L$ + VLLs}}}} \\ [0.5ex] 
 \cline{2-6}
   \multicolumn{1}{|c|}{} & \rule{0pt}{3.0ex} ${l_i}_\L = ({\nu_i}_\L,{e_i}_\L)^T $ & $\mathbf{1}$ & $\mathbf{2}$ & $-1/2$ & $-1$ & \multicolumn{1}{|c|}{} \\ [0.5ex] 
 \cline{2-6}
 \multicolumn{1}{|c|}{} & \rule{0pt}{3.0ex} ${e_i}_\R$ & $\mathbf{1}$ & $\mathbf{1}$ & $-1$ & $-1$ & \multicolumn{1}{|c|}{} \\ [0.5ex] 
\cline{2-6}
 \multicolumn{1}{|c|}{} & \rule{0pt}{3.0ex} ${N_i}_\R$ & $\mathbf{1}$ & $\mathbf{1}$ & $0$ & $-1$ & \multicolumn{1}{|c|}{} \\  [0.5ex] 
 \cline{2-6}
 \multicolumn{1}{|c|}{} & \rule{0pt}{3.0ex} $\Phi$ & $\mathbf{1}$ & $\mathbf{2}$ & $1/2$ & $0$ & \multicolumn{1}{|c|}{} \\ [0.5ex] 
 \cline{2-6}
 \multicolumn{1}{|c|}{} & \rule{0pt}{3.0ex} $S$ & $\mathbf{1}$ & $\mathbf{1}$ & $0$ & $2$ & \multicolumn{1}{|c|}{} \\ [0.5ex] 
\cline{1-6}
 \multicolumn{1}{c|}{} & \rule{0pt}{3.0ex} ${\ell_i}_{\L,\R} = ({\mathcal{N}_i}_{\L,\R},{\mathcal{E}_i}_{\L,\R})^T$ & $\mathbf{1}$ & $\mathbf{2}$ & $-1/2$ & $1$ & \multicolumn{1}{|c|}{} \\ [0.5ex] 
\cline{2-6}
 \multicolumn{1}{c|}{} & \rule{0pt}{3.0ex} ${E_i}_{\L,\R}$ & $\mathbf{1}$ & $\mathbf{1}$ & $-1$ & $1$ & \multicolumn{1}{|c|}{} \\[0.5ex]
 \cline{2-7}
\end{tabular}
    \caption{Field content and respective charges under the gauge group, for both the vanilla $B-L$ model and its extension with VLLs as first presented in~\cite{Feng:2016ysn}. The index $i$ runs over the three generations of fermions.}
    \label{tab:field_content}
\end{table}
Let us first consider the vanilla gauged $U(1)_{B-L}$ model in which the SM is extended with a singlet neutral lepton $N_\R$ for each generation, to ensure anomaly cancellation. Considering kinetic mixing effects (see details in Refs.~\cite{Hati:2020fzp,Bento:2023weq}), the following terms must be added to the SM covariant derivative $D_\mu^{\rm SM}$: 
\begin{equation}
D_\mu = D_\mu^{\rm SM}+ ie (\varepsilon \, Q + \varepsilon_{B-L} \, Q^{B-L}) \, Z'_\mu \, ,
\label{eq:covder}
\end{equation}
where $Q$ ($Q^{B-L}$) is the electric ($B-L$) charges and $Z'$ is a new neutral gauge boson. The parameters $\varepsilon$ and $\varepsilon_{B-L}$ can be expressed in terms of the kinetic mixing parameter and the $B-L$ gauge coupling, respectively. Kinetic mixing is constrained from $Z$-pole experiments~\cite{ALEPH:2005ab,CDF:2010zwq} to be small. Constraints on several extended neutral gauge structures were analysed in~\cite{Leike:1998wr,Erler:1999ub} and the new mixing angle is always constrained to be smaller than $10^{-3}$, approximately. In this limit, the $W$ and $Z$ boson masses are given by the usual SM expressions. A complex scalar singlet $S$ is introduced to break the $B-L$ symmetry spontaneously, generating a $Z^\prime$ mass through its vacuum expectation value (VEV). 
The field content of the model and corresponding transformation properties under the SM gauge group and $B-L$ charges are summarised in Table~\ref{tab:field_content}. 

The scalar potential invariant under the full gauge symmetry is given by
\begin{equation}
V_\text{Higgs} \,=\,\mu^2 |\Phi|^2 \,+\, m^2 |S|^2\,+\,\frac{\lambda_1}{2} |\Phi|^4 \,+\, 
\frac{\lambda_2}{2} |S|^4 \,+\, \lambda_3 |\Phi|^2 |S|^2 \,,    
\end{equation}
in which all parameters are real. After electroweak and $U(1)_{B-L}$ spontaneous symmetry breaking (SSB), the neutral component of the SM Higgs doublet $\Phi$ and $S$ acquire the real VEVs 
\begin{equation}
   \left< \phi^0 \right> = v/ \sqrt{2}\simeq 174~\GeV\;,\;\left< S \right> = w/ \sqrt{2}\,.
\end{equation}
In the limit of small kinetic mixing, the $Z^\prime$ mass stems predominantly from $\left< S \right>$ with
\begin{equation}
    m_{Z^\prime} \simeq 2 \, e \, |\varepsilon_{B-L}| w \, ,
    \label{eq:vevw}
\end{equation}
which should be about 17~MeV if the ATOMKI anomaly is to be explained by this $B-L$ vector boson. 

The remaining degrees of freedom mix and give rise to two CP-even scalars, $h$ and $\chi$, with the first being identified with the SM-like Higgs boson. The quartic couplings, which can be written in terms of the physical masses and the mixing angle $\alpha$,
\bea
\lambda_1 = \frac{1}{v^2}\left(c^2_\alpha \,m^2_h \,+\, s^2_\alpha\,m^2_\chi\right) \; ,\; 
\lambda_2 = \frac{1}{w^2}\left(s^2_\alpha \,m^2_h \,+\, c^2_\alpha\,m^2_\chi\right) \; , \;
\lambda_3 = \frac{1}{v\,w}\left(m^2_h \,-\, m^2_\chi\right)\,s_\alpha \,c_\alpha \,,
\label{eq:quart}
\eea
have to satisfy the following boundedness-from-below and unitarity constraints~\cite{Coimbra:2013qq}:
\be
0<\lambda_{1,2}\,\leq\,8\pi\;\;\; , \;\;\; 
-\sqrt{\lambda_1 \lambda_2} < \lambda_3\,\leq\,8\pi\;\;\; , \;\;\; 
|a_\pm|\,\leq\,8\pi\, ,
\label{eq:scalar_unitarity1}
\ee
with
\be
a_\pm \,=\, \frac{3}{2}\,\lambda_1\,+\,\lambda_2\,\pm\,
\sqrt{\left(\frac{3}{2}\,\lambda_1\,+\,\lambda_2\right)^2\,+\,\lambda_2^2}\,.
\label{eq:scalar_unitarity2}
\ee
The simplified notation $s_x (c_x) \equiv \sin x \left( \cos x \right)$ was employed. The coupling modifier of the Higgs-like particle to both electroweak gauge bosons and quarks is $c_\alpha$~\cite{Emam:2007dy}. This is not the case for leptons within an extended leptonic sector. Such quantities are well probed experimentally~\cite{CMS:2022dwd} and $c_\alpha$ has to be close to $1$ to comply with the expected SM behaviour for $h$. 

The strongest limit on the new-scalar mass $m_\chi$ comes from the unitary constraint $|a_{+}| \leq 8\pi$ -- see Eq.~(\ref{eq:scalar_unitarity1}) -- which leads to
\begin{equation}
    \dfrac{3}{2}\dfrac{m_h^2}{v^2} + \dfrac{m_\chi^2}{w^2} + \sqrt{ \left( \dfrac{3}{2}\dfrac{m_h^2}{v^2} + \dfrac{m_\chi^2}{w^2} \right)^2 + \left( \dfrac{m_\chi^2}{w^2} \right)^2 } \leq 8 \pi \quad , \quad w^{-2} = \dfrac{4 e^2 \varepsilon^2_{B-L}}{m^2_{Z^\prime}} \, ,
    \label{eq:limitMassChi}
\end{equation}
if one assumes $\lambda_1 \simeq m^2_h / v^2$ and $\lambda_2 \simeq m^2_\chi / w^2$. Taking the lowest possible value for $\varepsilon_{B-L}$ from~\cite{Hati:2020fzp}, the value $\varepsilon_{B-L}=2 \times 10^{-3}$ gives $m_\chi \lesssim 45~\text{GeV}$. Therefore the range one has to consider for the $B-L$ parameters implies that $m_\chi < m_h / 2$, and thus the decay channel $h \rightarrow \chi \chi$ is allowed, as well as $h \rightarrow Z^\prime Z^\prime$. Suppression of such decay modes further forces the limit $c_\alpha \simeq 1$, which is going to be used throughout the text. Before proceeding, let us briefly see the possible impact of not considering $c_\alpha$ to be $1$. Taking the value for the Higgs coupling modifier to the $Z$ boson, $\kappa_Z = 1.04 \pm 0.07$~\cite{CMS:2022dwd}, one can thus consider $c_\alpha \simeq 0.97$ instead of $c_\alpha \simeq 1$. In this case, for the same value of $\varepsilon_{B-L}$, the upper bound on the mass is even smaller, yielding $m_\chi \lesssim 34~\text{GeV}$.

The effective vector (V) and axial (A) couplings of $Z'$ with the neutron, proton and SM leptons can be cast in terms of $\varepsilon$ and $\varepsilon_{B-L}$ as shown in Table~\ref{tab:couplings}. We also present the experimental range for such quantities when $X(17)$ is the $B-L$ boson. Such limits were taken from~\cite{Hati:2020fzp}, with updated values in parenthesis from~\cite{Denton:2023gat}. The most stringent electron neutrino bounds come from recent coherent elastic neutrino nucleus scattering (CEvNS) experiments~\cite{Denton:2023gat}. Regarding muon neutrinos, the bounds are obtained following the detailed analysis done in~\cite{Hati:2020fzp}, in which CHARM-II data was used~\footnote{In~\cite{Denton:2023gat}, although the authors do not present bounds for muon or tau neutrinos, they comment that constraints on the coupling to muon neutrinos are only slightly less stringent than the ones to electron neutrinos. Regarding tau neutrinos, the coupling is about an order of magnitude less constrained~\cite{Kling:2020iar}.}.
\begin{table}[]
    \centering
   \begin{tabular}{||c | c | r||} 
 \hline
 \rule{0pt}{2.5ex}   
Couplings \, & \, Vanilla $B-L$ \, & \, Experimental range  \hspace{9em} \\ [0.5ex] 
 \hline\hline 
\rule{0pt}{2.5ex}   
 $\varepsilon_n^V$ & $\varepsilon_{B-L}$ & $ 2 \times 10^{-3} \left[ 4.1 \times 10^{-3} \right] \lesssim |\varepsilon_{n}^V| \lesssim 15 \times 10^{-3} \left[ 5.3 \times 10^{-3} \right]$ \hspace{0.01em} \hspace{0.9em} \\ [0.5ex] 
 \hline
 \rule{0pt}{2.5ex}  
 $\varepsilon_p^V$ & $ \varepsilon + \varepsilon_{B-L}$ &  $ \left[ 0.7 \times 10^{-3} \right] \lesssim |\varepsilon_{p}^V| \lesssim 1.2 \times 10^{-3} \left[ 1.9 \times 10^{-3} \right]$ \hspace{0.95em} \\ [0.5ex]  
 \hline
 \rule{0pt}{2.5ex}  
  $\varepsilon_{ee}^V$ & $-(\varepsilon+\varepsilon_{B-L})$ & \hspace{0.01em} $ 0.4 \times 10^{-3} \left[ 0.63 \times 10^{-3} \right] \lesssim |\varepsilon_{ee}^V| \lesssim 2 \times 10^{-3} \left[ 1.2 \times 10^{-3} \right] $ \hspace{0.4em} \hspace{1em} \\ [0.5ex]  
 \hline
 \rule{0pt}{2.5ex}
 $\varepsilon_{ee}^A$ & $0$ & $|\varepsilon_{ee}^A| \lesssim 2.6 \times 10^{-9}$ \hspace{9.35em} \\ [0.5ex] 
 \hline
 \rule{0pt}{2.5ex}  
  $\varepsilon_{\nu_e \nu_e}^A$ & $\varepsilon_{B-L}$ & $|\varepsilon_{\nu_e \nu_e}^A| \lesssim 1.2 \times 10^{-5} \left[ \left( 3.5 - 4.5 \right) \times 10^{-6} \right] $ \hspace{0.5em} \\ [0.5ex] 
 \hline
 \rule{0pt}{2.5ex}  
  $\varepsilon_{\nu_\mu \nu_\mu}^A$ & $\varepsilon_{B-L}$ & $|\varepsilon_{\nu_\mu \nu_\mu}^A| \lesssim 12.2 \times 10^{-5}$ \hspace{8.85em} \\ [0.5ex] 
 \hline
\end{tabular}
    \caption{Couplings of SM particles to the $Z^\prime$ boson in the vanilla $B-L$ model, \textit{i.e.} without VLLs, and the experimental ranges that explain the ATOMKI anomalies. Such bounds are taken from~\cite{Hati:2020fzp}, with updated values from~\cite{Denton:2023gat} in parenthesis. The lower indices label the following particles: $n$ for neutron, $p$ for proton, $e$ for electron, and $\nu_e$ ($\nu_\mu$) for the electron (muon) neutrino.}
    \label{tab:couplings}
\end{table}
The ``protophobic" nature of the $X(17)$ particle becomes clear from the experimental ranges. The right value for the couplings $\varepsilon_p^V$ and $\varepsilon_{ee}^V$ can be achieved in the vanilla $B-L$ scenario. However, this is not the case for $\varepsilon_n^V$ and $\varepsilon_{\nu \nu}^A$ which are predicted to be equal by the model but are required to be experimentally very different (see Table~\ref{tab:couplings}). The solution put forward in~\cite{Feng:2016ysn} consists in adding VLLs to the model so that the $Z^\prime$ couplings with neutrinos are affected, leaving those with the neutron and the proton unaltered. We will put this solution to the test in what follows.

Let us then consider the vanilla $B-L$ model extended with three generations of VLL doublets ${\ell_i}_{\L,\R}$ and singlets ${E_i}_{\L,\R}$ with $Q_{B-L}$ charges given in Table~\ref{tab:field_content}. The doublets ${l_i}_\L$ and singlets ${e_i}_\R$ are the usual SM fields. The lepton mass and Yukawa Lagrangian is:
\begin{equation}
\begin{split}
    \mathcal{L}_\text{lepton} =& -y_l^{ij} \, \overline{l_i}_\L \Phi \, {e_j}_\R - y_\nu^{ij} \,  \, \overline{l_i}_\L \tilde{\Phi} \, {N_j}_\R - \dfrac{1}{2} y_M^{ij} S \, \overline{N_i}_\R^{c} {N_j}_\R - \lambda_\L^{ij} S^{*} \, \overline{l_i}_\L {\ell_j}_\R - \lambda_{\rm E}^{ij} S \, \overline{E_i}_\L {e_j}_\R \\
& - M_\L^{ij}\overline{\ell_i}_\L {\ell_j}_\R - M_{\rm E}^{ij}\overline{E_i}_\L {E_j}_\R - h^{ij} \, \overline{\ell_i}_\L \Phi \, {E_j}_\R - k^{ij}  \, \overline{E_i}_\L \Phi^\dagger \, {\ell_j}_\R \, + \text{H.c.} \; ,\; i,j=1,2,3\,.
\end{split}
\label{eq:leptonLagrangian}
\end{equation}
in which $y$, $\lambda$, $h$ and $k$ are $3 \times 3$ complex Yukawa coupling matrices and $M_{\rm L,E}$ are 
$3 \times 3$ bare-mass matrix terms. We highlight the fact that SM fields only mix with new leptons that carry the same SM quantum numbers through $\overline{l}_\L \ell_\R$ and $\overline{E}_\L e_\R$. For simplicity, we will assume that the Yukawa couplings $y_l$, $\lambda_{\rm L,E}$, $h$ and $k$, as well as the mass terms $M_{\rm L,E}$, are real and diagonal. This implies that the SM fields of a given generation will only mix with the vector-like fields of the same generation, preventing charged lepton flavour violation processes from occurring. After SSB, the charged-lepton mass terms are:
\begin{equation}
\mathcal{L}_\text{c.l.}=\overline{\psi^\ell_\L} \, \mathcal{M}_\ell \, \psi^\ell_\R + \text{H.c.} \;,\;     \psi^\ell_{\L,\R} = \left( e_{\L,\R} \, , \, \mathcal{E}_{\L,\R} \, , \, E_{\L,\R} \right)^T\;,\;
\mathcal{M}_\ell = 
    \left(
    \setlength{\tabcolsep}{10pt} 
\renewcommand{\arraystretch}{1.5} 
 \begin{array}{ccc} 
y_l \, \dfrac{v}{\sqrt{2}} & \lambda_\L \, \dfrac{w}{\sqrt{2}} & 0 \\[0.25cm]
0 & M_\L & h \, \dfrac{v}{\sqrt{2}} \\[0.25cm]
\lambda_{\rm E} \, \dfrac{w}{\sqrt{2}} & k \, \dfrac{v}{\sqrt{2}} & M_{\rm E}
\end{array}
\right) \, .
\label{eq:Mell}
\end{equation}
Each entry in the $9 \times 9$ matrix $\mathcal{M}_\ell$ should be understood as a $3 \, \times \, 3$ diagonal block in generation space with distinct $y_l^i$, $\lambda_{\rm L,E}^i$, $M_{\rm L,E}^i$, $k_i$ and $h_i$ elements in the diagonal ($i=1,2,3$). The full mass matrix can be diagonalised by a bi-unitary rotation, $\mathcal{M}_\ell^\text{diag} = U_\L^{\dagger} \, \mathcal{M}_\ell \, U_\R$, the three lightest eigenstates being the usual SM charged leptons. In the limit of no lepton-generation mixing, each family can be treated separately by diagonalising its corresponding $3\times 3$ mass matrix $\mathcal{M}_\ell^i$ which has the form of $\mathcal{M}_\ell$ with the matrix blocks replaced by the generation coupling or mass parameter.

For the neutral lepton mass terms we have
\begin{equation}
\mathcal{L}_\text{n.l.}=
{\overline{{\psi^\nu_\L}^c}} \, \mathcal{M}_\nu \, \psi_\L^\nu + \text{H.c.}\;,\;\psi_\L^\nu = \left( \nu \, , \, N^c \, , \, \mathcal{N} \, , \, \mathcal{N}^c \right)_\L^T\;,\;
    \mathcal{M}_\nu = 
    \left(
    \setlength{\tabcolsep}{10pt} 
\renewcommand{\arraystretch}{1.5} 
 \begin{array}{cccc} 
0 \, & \, y_\nu \dfrac{v}{\sqrt{2}} \, & \, 0 \, & \, \lambda_\L \dfrac{w}{\sqrt{2}} \\[0.3cm]
y_\nu \dfrac{v}{\sqrt{2}} \, & \, y_M \dfrac{w}{\sqrt{2}} \, & \, 0 \, & \, 0 \\[0.3cm]
0 \, & \, 0 \, & \, 0 \, & \, M_\L \\[0.3cm]
\lambda_\L \dfrac{w}{\sqrt{2}} \, & \, 0 \, & \, M_\L \, & \, 0
\end{array}
\right) \, ,
\label{eq:Mnu}
\end{equation}
where $\mathcal{M}_\nu $ is now a $12 \times 12$ symmetric matrix, diagonalised by a single unitary matrix $U_\nu$ as $\mathcal{M}_\nu^\text{diag}=U_\nu^T \mathcal{M}_\nu U_\nu $. We expect the light active neutrino masses to be generated via a type-I-like seesaw mechanism, relying on the Majorana mass terms of the singlet right-handed neutrinos, with some contribution from the vector-like neutral fields. As previously mentioned, we work under the assumption that all couplings are diagonal in generation space. However, this is not the case for the couplings $y_\nu$, since this sector must reproduce the PMNS matrix, from which leptonic mixing arises. This fact increases the complexity of the neutral leptonic sector, namely one should indeed work with a $12 \, \times \, 12$ matrix instead of considering a schematic $4 \, \times \, 4$ approach for each generation if leptonic mixing effects are relevant. 

After defining the transformations which bring leptons to their mass eigenstate basis, we can determine the $Z^\prime$ couplings with quarks and leptons stemming from the fermionic kinetic terms $\overline{f} \slashed{D} f$, taking the covariant derivative in Eq.~(\ref{eq:covder}). We write the generic vector and axial $Z^\prime$-fermion couplings $\varepsilon_{ab}^{V(A)}$ as
\begin{equation}
    \mathcal{L} \supset e {\overline{f_a}} \gamma^\mu \left( \varepsilon^V_{ab} + \gamma^5 \varepsilon^A_{ab} \right) {f}_b \, Z_\mu^\prime \,.
\end{equation}
For the up- and down-type quarks, the axial $Z^\prime$ coupling vanishes, while the vector ones are $\varepsilon^V_{qq}=\varepsilon Q_q + \varepsilon_{B-L} Q_q^{B-L}$, for $q=u,d$. From this, it is straightforward to obtain the vector couplings to protons ($p$) and neutrons ($n$):
\begin{equation}
    \varepsilon^V_p = \varepsilon + \varepsilon_{B-L} \quad , \quad \varepsilon^V_n = \varepsilon_{B-L} \, .
\end{equation}
As expected, $\varepsilon^V_{p,n}$ are the same as in the vanilla $B-L$ model (see Table~\ref{tab:couplings}) since no new coloured particles have been added. This is obviously not the case for leptons due to the presence of VLLs.

In the VLL-extended $B-L$ model the vector and axial $Z'$ couplings to the charged leptons $\ell^\pm_{a}\ell^\mp_{b}$ can be written as 
\begin{equation}
\varepsilon^{V(A)}_{\ell^\pm_{a}\ell^\mp_{b}} = \dfrac{1}{2} \sum_{i=1}^3 \left( \varepsilon \, Q_i + \varepsilon_{B-L} \, Q_i^{B-L} \right) \left[ \left( U_{\R}^\dagger \right)_{ai} \left( U_{\R} \right)_{ib} \pm \left( U_{\L}^\dagger \right)_{ai} \left( U_{\L} \right)_{ib} \right] \, ,
  \label{eq:LRcharged}
\end{equation}
with $Q_i = \{ -1,-1,-1 \} $ and $Q_i^{B-L} = \{ -1,+1,+1\} $. The indices $a,b=1,2,3$ identify the corresponding leptonic mass eigenstates $\ell^\pm$, within the same generation. Without VLLs, the vector $Z'$ coupling to the SM charged leptons is simply given by $\varepsilon_{\ell^\pm_{1}\ell^\mp_{1}}^V = - (\varepsilon + \varepsilon_{B-L})$, while the axial one vanishes. The VLLs will change these couplings since, within each generation, they mix with the SM charged leptons. For the neutral leptons, we have instead
\begin{equation}
    \varepsilon^{V(A)}_{\nu_a \nu_b} = \pm \dfrac{1}{2} \sum_{i=1}^{12} \, \varepsilon_{B-L} \, Q_i^{B-L} \left[ \left(U_\nu^{*}\right)_{ia} \left( U_\nu \right)_{ib} \mp \left(U_\nu\right)_{ia} \left( U_\nu^{*} \right)_{ib}  \right] \, ,
    \label{eq:VAneutrino}
\end{equation}
with each entry of $Q_i^{B-L}= \{ -1,-1,+1,+1\}$ accounting for the three generations of each neutral field. In the absence of VLLs, the couplings of the lightest mass eigenstate are $\varepsilon_{\nu_1 \nu_1}^V=0$ and $\varepsilon_{\nu_1 \nu_1}^A=\varepsilon_{B-L}$, which is in agreement with~\cite{Hati:2020fzp}.  

Up to this point, one should clarify why it is possible to make a direct correspondence between the $Z^\prime$ coupling to the lightest neutral mass eigenstate with the one to a definite flavour. The bounds on neutrino couplings come from scattering experiments and effects of neutrino masses and mixing on elastic neutrino-electron scattering have been analysed in~\cite{Grimus:1997aa}. The authors concluded that for neutrino masses of eV order, the interference cross section is approximately six orders of magnitude suppressed compared to the pure weak and electromagnetic cross sections for MeV neutrinos. Thus, effects coming from leptonic mixing are not expected to be significant. Therefore, one can claim that the effective $Z^\prime$ couplings to neutrinos are identical to that of the neutron (up to a global sign), without VLLs. Neglecting neutrino masses also works as a good approximation and we will take this limit from now on.

\section{Can $\varepsilon_{\nu\nu}$ be naturally (and consistently) suppressed?}

As shown in Eqs.~\eqref{eq:LRcharged} and \eqref{eq:VAneutrino}, the $Z'$ couplings to leptons are modified due to the presence of VLLs. Next, we will investigate whether the suppression of $\varepsilon_{\nu\nu}$ required to explain the ATOMKI anomalies is achievable taking into account other constraints. In order to check this, we start by performing some analytical estimates based on reasonable assumptions regarding the VLL/neutrino couplings and mass terms. It is convenient to define the parameters
\begin{equation}
    \Lambda_{\L,\rm E} \equiv \dfrac{\lambda_{\L,\rm E} \, w}{\sqrt{2} \, M_{\L,\rm E}} \, ,
    \label{eq:mixing_parameters}
\end{equation}
which control the VLL mixing with SM leptons. In the following, we present the form for the matrices $U_{\L,\R}$ and $U_\nu$. Next, we show why $Z$-pole measurements do not constrain the mixing with the SM, and proceed to explore the scenarios put forward in~\cite{Feng:2016ysn} with large mixing and $Q_{\text{VLL}}^{B-L}=1$, and in~\cite{Denton:2023gat} where small mixing and large $B-L$ charges were considered.

We assume that the $y_\nu$-term in the mass matrix of Eq.~(\ref{eq:Mnu}) can be neglected since, in the present framework, we naturally expect $\frac{y_\nu v}{y_M w},\frac{y_\nu v}{\lambda_\L w }, \frac{y_\nu v}{M_\L} \ll 1$, in order to have naturally suppressed neutrino masses\footnote{A natural approach to such seesaw-type mass matrices also implies that $\lambda_\L w \ll M_\L$, but we do not require small mixing \textit{a priori}.}~\cite{Grimus:2000vj}. Also, as we have seen, effects from neutrino masses and mixing are not significant in scattering experiments, from which the relevant couplings are constrained. Thus, under these premises, we can follow the same approach as for charged leptons, \textit{i.e.} diagonalise the $4\times 4$ neutral lepton mass matrix for each generation. This leads to the following $U_\nu$ rotation matrix: 
\begin{equation}
    U_\nu = 
    \left(
    \setlength{\tabcolsep}{12pt} 
\renewcommand{\arraystretch}{1.5} 
 \begin{array}{cccc} 
-\dfrac{1}{\sqrt{1+\Lambda_{\L}^2}} & 0 & -\dfrac{1}{\sqrt{2}} \dfrac{\Lambda_{\L}}{\sqrt{1+\Lambda_{\L}^2}} &  \dfrac{1}{\sqrt{2}} \dfrac{\Lambda_{\L}}{\sqrt{1+\Lambda_{\L}^2}} \\[0.4cm]
0 & 1 & 0 & 0 \\[0.4cm]
\dfrac{\Lambda_{\L}}{\sqrt{1+\Lambda_{\L}^2}} & 0 & -\dfrac{1}{\sqrt{2}} \dfrac{1}{\sqrt{1+\Lambda_{\L}^2}} &  \dfrac{1}{\sqrt{2}} \dfrac{1}{\sqrt{1+\Lambda_{\L}^2}} \\[0.4cm]
0 & 0 & \dfrac{1}{\sqrt{2}} & \dfrac{1}{\sqrt{2}}
\end{array}
\right) \, .
\end{equation}

Regarding the charged sector, and as in Ref.~\cite{Feng:2016ysn}, we take the simple case of vanishing Yukawa couplings $h=k=0$ in Eq.~\eqref{eq:Mell} which leads to no mixing between the vector-like fields of each generation\footnote{Despite the fact that mixing between VLLs and SM leptons is not constrained by precision measurements, this does not happen for the mixing between the VLLs. Such mixing is constrained by the oblique parameters, making the no-mixing limit well motivated.}. Still, the SM charged-lepton fields mix with the VLLs via the $\lambda_{\L,\rm E} \, w$ terms. We start by neglecting the term $y_l$ of Eq.~(\ref{eq:Mell}), since $\frac{y_l v}{\lambda_{\L,\rm E} w }, \frac{y_l v}{M_{\L,\rm E}} \ll 1$. Under such conditions, the mass matrix is in fact diagonalised with just two mixing angles. The matrices $U_{\L,\R}$ can be simply written in the following form:
\begin{equation}
    U_\L = 
    \left(
    \setlength{\tabcolsep}{10pt} 
\renewcommand{\arraystretch}{1.5} 
 \begin{array}{ccc} 
\dfrac{1}{\sqrt{1+\Lambda_{\L}^2}} & \dfrac{\Lambda_{\L}}{\sqrt{1+\Lambda_{\L}^2}} & 0 \\[0.3cm]
-\dfrac{\Lambda_{\L}}{\sqrt{1+\Lambda_{\L}^2}} & \dfrac{1}{\sqrt{1+\Lambda_{\L}^2}} & 0 \\[0.3cm]
0 & 0 & 1
\end{array}
\right) 
\quad , \quad 
 U_\R = 
    \left(
    \setlength{\tabcolsep}{10pt} 
\renewcommand{\arraystretch}{1.5} 
 \begin{array}{ccc} 
\dfrac{1}{\sqrt{1+\Lambda_{\rm E}^2}} & 0 & \dfrac{\Lambda_{\rm E}}{\sqrt{1+\Lambda_{\rm E}^2}} \\[0.3cm]
0 & 1 & 0 \\[0.3cm]
-\dfrac{\Lambda_{\rm E}}{\sqrt{1+\Lambda_{\rm E}^2}} & 0 & \dfrac{1}{\sqrt{1+\Lambda_{\rm E}^2}} \\[0.3cm]
\end{array}
\right) \, .
\label{eq:ULUR}
\end{equation}
The left- and right-handed couplings of the SM charged leptons $\ell^\pm_1$ with the $Z$ boson are~\cite{Dermisek:2013gta}:
\begin{equation}
    g_{\L,\R}^{Z \ell^\pm_1 \ell^\mp_1} = \dfrac{g}{c_W} \sum_{i=1}^3 \left( T^3_i - s_W^2 Q_i \right) \left( U_{\L,\R}^\dagger \right)_{1i} \left( U_{\L,\R} \right)_{i1} \, ,
\end{equation}
with $T^3$ being the associated weak isospin, and $\theta_W$ is the weak mixing angle. Since the non-vanishing entries of the first columns of $U_{\L,\R}$ connect fields with the same weak isospin and electric charge, the $Z$ couplings are automatically satisfied and $Z$-pole constraints do not restrict the mixing. Solely considering this, the mixing can be general \textit{a priori}.

\subsection{Large mixing and $Q_\text{VLL}^{B-L}=1$}

In Ref.~\cite{Feng:2016ysn}, a $B-L$ scenario extended with VLLs has been proposed to account for the suppression of $\varepsilon_{\nu \nu}$, with the corresponding charges given in Table~\ref{tab:field_content}. From Eq.~(\ref{eq:VAneutrino}), it is straightforward to obtain:
\begin{equation}
    \varepsilon^A_{\nu_1 \nu_1} = \varepsilon_{B-L} \left( \dfrac{1-\Lambda_{\L}^2}{1+\Lambda_{\L}^2} \right) \, .
\end{equation}
Although this coupling is still subjected to corrections of the order of the small neutrino parameters we have neglected, the approximation turns out to be rather good. It is clear that $\Lambda_\L \simeq 1$ is required to achieve the suppression of $\varepsilon^A_{\nu_1 \nu_1}$ with respect to $\varepsilon_n^{V}=\varepsilon_{B-L}$. Defining $\Lambda_\L = 1-\delta_\L$ with $\delta_\L \ll 1$, we have  
\begin{equation}
    \dfrac{\varepsilon^A_{\nu_1 \nu_1}}{\varepsilon_{B-L}} =  \delta_\L + \mathcal{O}\left( \delta_\L^2 \right) \ll 1 \, ,
\end{equation}
which reflects what we have just said. At this point, one could already object against the fact that the fine tuning required between $\lambda_\L w$ and $M_\L$ to ensure $\Lambda_\L \simeq 1$ is completely unjustified in the model, since those parameters are rather arbitrary. Nevertheless, we will turn a blind eye to this detail and proceed with the analysis.

In one assumes $\Lambda_\L \simeq 1$, then the coupling $\varepsilon_{\ell_1^\pm \ell_1^\mp}^A$ reads
\begin{equation}
    \varepsilon_{\ell_1^\pm \ell_1^\mp}^A \simeq \dfrac{1}{2} \, \varepsilon_{B-L} \left( \dfrac{\Lambda_{\rm E}^2 - 1}{1+\Lambda_{\rm E}^2} \right) \, .
\end{equation}
From the values of Table~\ref{tab:couplings}, the axial coupling to the first generation, $\varepsilon_{ee}^A$, must be strongly suppressed, even more than the coupling to neutrinos with respect to $\varepsilon_{B-L}$. Considerations from~\cite{Feng:2016jff,Feng:2016ysn} solidly support that the axial couplings with the remaining generations must also be suppressed. Thus, the condition $\Lambda_{\rm E} \simeq 1$ is also required. Considering $\Lambda_{\rm E} = 1 - \delta_{\rm E}$ with $\delta_{\rm E} \ll 1$, then
\begin{equation}
    \dfrac{\varepsilon_{\ell_1^\pm \ell_1^\mp}^A}{\varepsilon_{B-L}} \simeq - \dfrac{1}{2} \delta_{\rm E} + \mathcal{O}\left( \delta_{\rm E}^2 \right) \ll 1 \, .
\end{equation}
Considering $\Lambda_{\L}\simeq 1$, required to suppress $\varepsilon_{\nu\nu}$, also implies $\Lambda_{\rm E}\simeq 1$ in order to keep $\varepsilon_{\ell_1^\pm \ell_1^\mp}^A$ axial coupling small~\cite{Hati:2020fzp}.

Let us now see how such large mixing impacts the Higgs sector. Given that the SM leptons mix with the VLLs, the coupling of the SM charged leptons with the Higgs-like particle $h$, stemming mostly from the scalar doublet, is given by
\begin{equation}
    - \mathcal{L} \supset \dfrac{m_\ell}{v} \, \kappa_\ell \, h \, \overline{\ell^\pm_\L} \ell^\pm_\R + \text{H.c.} \;.
\end{equation}
Taking into account the rotations in Eq.~\eqref{eq:ULUR} and since, in the present case, $y_l=(\sqrt{2} \, m_\ell / v) \left(U_\L \right)_{11} \left(U_\R^\dagger \right)_{11}$, the coupling modifier $\kappa_\ell$ is given by\footnote{In fact, the coupling modifier should also include the factor $c_\alpha$ that accounts for scalar mixing, which would even decrease the value of $\kappa_\ell$. In this approximation, we take $c_\alpha \simeq 1$.}
\begin{equation}
    \kappa_\ell = \left|\left(U_\L \right)_{11}\right|^2 \left|\left(U_\R\right)_{11}\right|^2  = \dfrac{1}{1+\Lambda_\L^2} \, \dfrac{1}{1+\Lambda_{\rm E}^2} \, .
\label{eq:kappal}    
\end{equation}
For $\Lambda_{\L,\rm E} = 1 - \delta_{\L,\rm E}$, with $\delta_{\L,\rm E} \ll 1$ as in~\cite{Feng:2016ysn,Hati:2020fzp}, we have
\begin{equation}
    \kappa_\ell = \dfrac{1}{4} + \mathcal{O}\left( \delta_\L, \delta_{\rm E} \right) \, ,
\end{equation}
which means that the couplings to the Higgs boson are just $25 \, \%$ of their SM value, which is in tension with what has been observed at the LHC, for instance, regarding the second generation, $\kappa_{\mu}=1.12^{+0.21}_{-0.22}$~\cite{CMS:2022dwd}. Despite the lack of experimental values on Table~\ref{tab:couplings} for the third generation, one would expect a similar disagreement with $\kappa_{\tau}=0.92 \pm 0.08$~\cite{CMS:2022dwd}.

Although $Z$-pole measurements allow for the large lepton mixing required to suppress $\varepsilon_{\nu\nu}$, the couplings of the SM charged leptons to the Higgs boson are significantly modified with respect to their SM values. Moreover, the authors of Ref.~\cite{Denton:2023gat} noticed that non-unitarity effects due to active-neutrino mixing with the extra neutral particles are unacceptably large in this case. Considering the first generation only, these effects can be parameterised by the parameter
\begin{equation}
     \delta_{e}=1 - |\left( U_\nu \right)_{11}|^2 \lesssim 0.04 \quad \Rightarrow \quad \Lambda_\L^2 \lesssim 1/24 \, ,
     \label{eq:limitLambdaL}
 \end{equation}
where the lower limit on $\delta_{e}$ comes from the strongest solar neutrino bound that directly contains the electron neutrino row normalization, which is the ${}^7\text{Be}$ measurement. This $90\%$ C.L. deviation obtained by~\cite{Denton:2023gat} uses the measurement from KamLAND and the theory prediction of the flux~\cite{Bergstrom:2016cbh}. This is another clear and strong indication that the large mixing required by neutrino coupling suppression appears to be in conflict with experimental evidences.

\begin{figure}
  \centering
  \includegraphics[height=7.0cm]{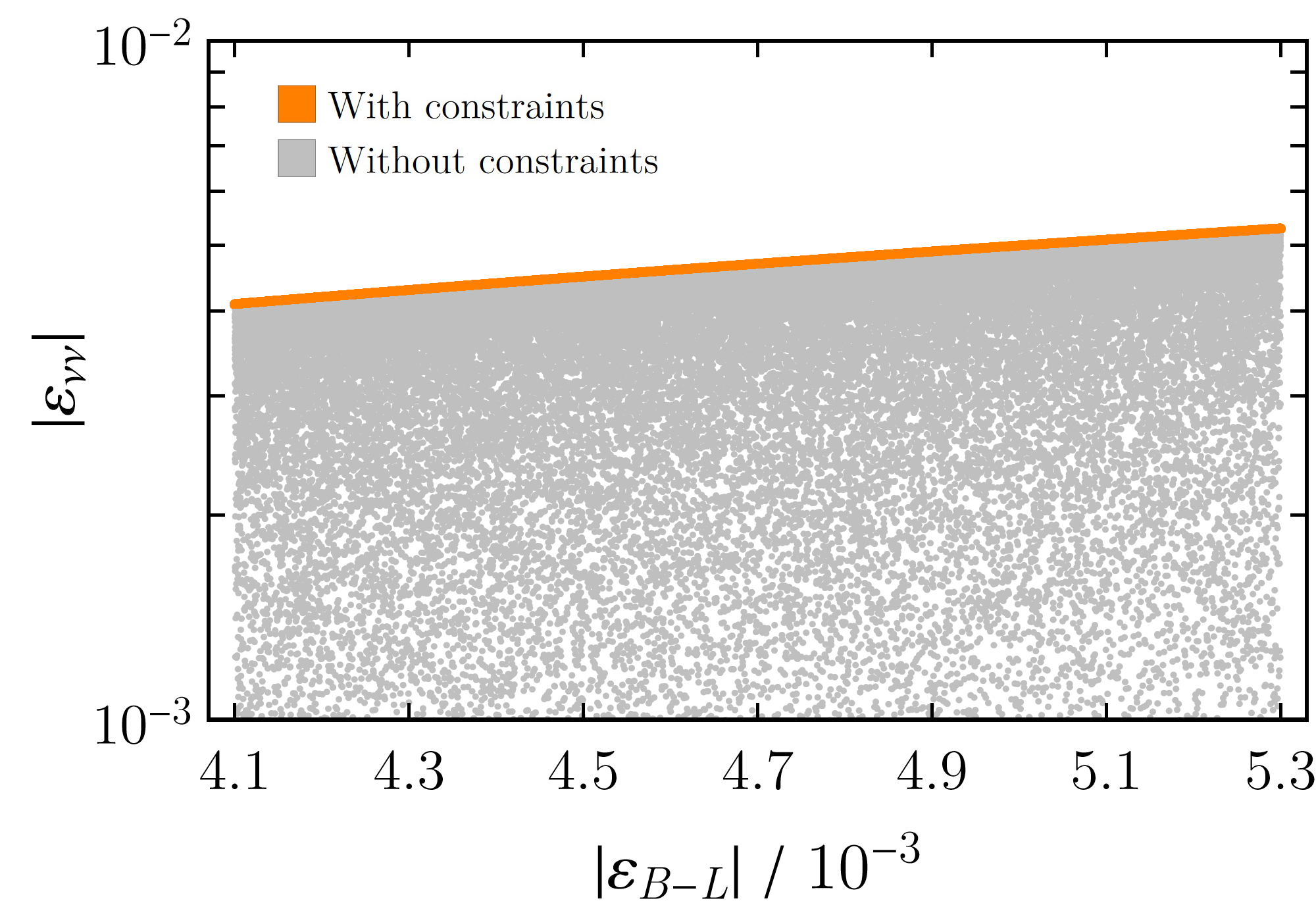}
  \caption{Numerical results on the plane $(|\varepsilon_{B-L}|,|\varepsilon_{\nu \nu}|)$ for the first generation, considering $Q_{B-L}^\text{VLL}=1$. Taking into account several constraints (see text for more details), neutrino-coupling suppression is not possible and $|\varepsilon_{\nu \nu}|=|\varepsilon_{B-L}|$, in orange. Without such constraints, the considered lepton mixing is general and suppression down to the desired values can be achieved: $|\varepsilon_{\nu \nu}|<|\varepsilon_{B-L}|$, in gray. Points with $|\varepsilon_{\nu \nu}| < 10^{-3}$ are not displayed for the sake of clarity of the figure.}
  \label{fig:suppression}
\end{figure}

Despite the fact that the approximations put forward so far are well motivated, we performed a numerical scan of the full model. We considered mixing between the vector-like fields and allowed $c_\alpha$ to vary. We imposed perturbativity taking all Yukawa couplings to be less than $\sqrt{4 \pi}$, $Z$-pole~\cite{Dermisek:2013gta} and $STU$ constraints~\cite{Bento:2023weq,Grimus:2008oblique,Lavoura:1993oblique}. Limits from Higgs invisible decay~\cite{ATLAS:2023tkt} and coupling modifiers~\cite{CMS:2022dwd} were also considered as well as neutrino non-unitarity constraints on the first generation, as previously described. The expected $X(17)$ behaviour was also checked through its couplings' values of Table~\ref{tab:couplings}. We allowed the heavy-lepton masses to range from $100~\text{GeV}$ up to $1~\text{TeV}$~\cite{Dermisek:2021ajd} and the new-scalar mass from $10$ to $62.5~\text{GeV}$. The mixing angles between the leptonic fields were also randomly chosen between $0$ and $\pi/2$. Our results on the plane $\left( |\varepsilon_{B-L}|, |\varepsilon_{\nu \nu}| \right)$ for the first generation are shown in Fig.~\ref{fig:suppression}. Considering the complete set of constraints just described, we found out that neutrino coupling suppression is indeed not viable within a large-mixing framework, as expected. That is, we could not obtain $|\varepsilon_{\nu \nu}|$ for the first generation that complies with the allowed values of Table~\ref{tab:couplings}. Without taking them into account, which allows the mixing to be general, we observed that $|\varepsilon_{\nu \nu}|$ could be as suppressed as needed. For the sake of clarity we do not present such low values in Fig.~\ref{fig:suppression}.

To summarise, the large lepton mixing required to suppress $\varepsilon_{\nu\nu}$ is in clear tension with experimental data, if the $B-L$ boson is to be the $X(17)$ particle. We showed that the couplings of the SM charged leptons to the Higgs boson are significantly modified with respect to their SM values. The authors of Ref.~\cite{Denton:2023gat} also pointed out that non-unitarity constraints on neutrino mixing impose strong limits on
the model as well. 
To circumvent this, the solution put forward in~\cite{Denton:2023gat} is to consider VLL fields with large $B-L$ charges, instead of $Q_\text{VLL}^{B-L}=1$, in order to achieve $\varepsilon_{\nu\nu}$ with small heavy-light neutrino mixing simultaneously. In the next section we show that in this case the dominant $Z^\prime$ contributions to the muon's anomalous magnetic moment are too large.

\subsection{Small mixing and large $B-L$ charges}

Let us now consider the scenario in which the $B-L$ charges $Q_\text{VLL}^{B-L}$ of ${\ell_i}_{\L,\R}, {E_i}_{\L,\R}$ are general. The new Higgs scalar $S$ must have charge equal to $Q_\text{VLL}^{B-L}+1$ for $\lambda_{\L,\rm E}$ to be nonzero as may be seen from Eq.~(\ref{eq:leptonLagrangian}). 
Note that different $B-L$ charges per generation would require extra scalar fields. Furthermore, now the field $S$ no longer provides a Majorana mass term to the right-handed neutrinos, and the ensuing type-I-like seesaw mechanism would be absent. Thus, the mass of active neutrinos must be generated otherways, for instance, by adding a third Higgs field. However, since working with massless neutrinos is a good approximation, this will not be crucial here. Using again Eq.~(\ref{eq:VAneutrino}), we now obtain:
\begin{equation}
    \varepsilon^A_{\nu_1 \nu_1} = \varepsilon_{B-L} \left( \dfrac{1-Q_\text{VLL}^{B-L}\Lambda_{\L}^2}{1+\Lambda_{\L}^2} \right) \, ,
\end{equation}
which is now suppressed if $Q_\text{VLL}^{B-L}\Lambda_\L^2 \simeq 1$. This implies that small lepton mixing (or small $\Lambda_\L$) requires large $B-L$ charges. 
By considering the limit on $\Lambda_\L$ derived from neutrino unitarity shown in Eq.~(\ref{eq:limitLambdaL}) together with imposing $Q_\text{VLL}^{B-L}\Lambda_\L^2 \simeq 1$, the inequality implies a lower bound on the $B-L$ charge of approximately $24$, as obtained in Ref.~\cite{Denton:2023gat}\footnote{Different analyses can lead to stronger constraints on $\delta_e$. As an example, in~\cite{Hu:2020oba}, non-unitarity is constrained to be less than $0.003$, which would give even larger $B-L$ charges.}.

Let us now assess how natural a $B-L$ charge of $24$ is in fact. Directly from the suppression condition $Q_\text{VLL}^{B-L}\Lambda_\L^2 \simeq 1$ together with Eqs.~(\ref{eq:mixing_parameters}) and~(\ref{eq:vevw}), we can write:
\begin{equation}
    Q_\text{VLL}^{B-L} \simeq \dfrac{1}{\Lambda_\L^2} \simeq \left( 57 - 95 \right) \left( \dfrac{M_\L}{130~\text{GeV}} \right)^2 \left( \dfrac{\sqrt{4 \pi}}{\lambda_\L} \right)^2 \, ,
\end{equation}
where the lower (upper) limit presented in parenthesis is obtained for the minimum (maximum) value of $\varepsilon_{B-L}$. We see that for the perturbativity-limit value $\lambda_\L \simeq \sqrt{4 \pi}$, and considering the conservative reference value of $M_\L = 130~\text{GeV}$ from Refs.~\cite{Feng:2016ysn,Denton:2023gat}, the lowest value for the charge is, in fact, $Q_\text{VLL}^{B-L} \simeq 57$. If one considers the lower bound of $800~\text{GeV}$ as in~\cite{Dermisek:2021ajd}, the minimum value for the charge turns out to be $2159$ (!). In~\cite{Chun:2020uzw}, the authors consider $500~\text{GeV}$, which would still yield a value much larger than $24$, namely $843$. However if one decides to turn a blind eye to the perturbativity limit of $\sqrt{4 \pi}$, and also assumes $M_\L \simeq 130~\text{GeV}$, then a $B-L$ charge of $24$ would still be possible if $\lambda_\L \simeq 5.46~(7.05)$, considering the lowest (largest) value of $\varepsilon_{B-L}$. Despite the fact that $Q_\text{VLL}^{B-L}=24$ is in clear high tension with perturbativity, for the remaining of our discussion we will proceed with this as the reference value for $Q_\text{VLL}^{B-L}$.

\begin{figure}
  \centering
  \includegraphics[height=8.0cm]{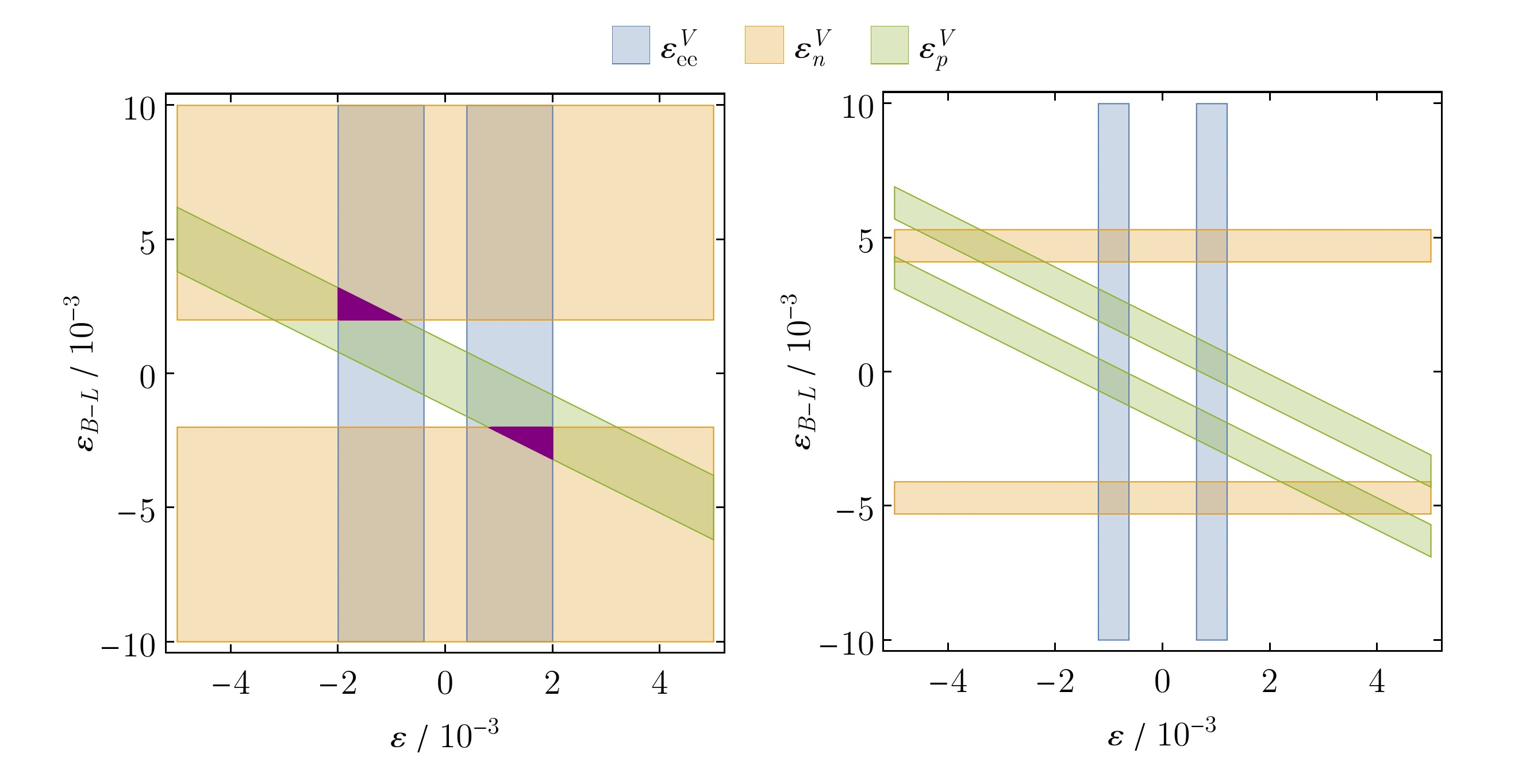}
  \caption{Limits on the couplings $\varepsilon$ and $\varepsilon_{B-L}$, when $Q_\text{VLL}^{B-L} \Lambda_{\L,\rm E}^2 \simeq 1$. The blue (orange) shaded band corresponds to the experimental range for $\varepsilon_{ee}^V$ ($\varepsilon_n^V$), while the green one represents the constraints on $\varepsilon_p^V=\varepsilon+\varepsilon_{B-L}$. On the left (right) plot, we have used the non-updated (updated) values from Table~\ref{tab:couplings}. On the left, there is a small overlap (in purple) of the three regions while this is not the case for the right panel. See text for more details.}
  \label{fig:limits_on_couplings}
\end{figure}

The relation $Q_\text{VLL}^{B-L} \Lambda_\L^2 \simeq 1$ has a direct impact on the $Z^\prime$ couplings with the SM charged leptons. Namely, once again, for the axial couplings we have
\begin{equation}
    \varepsilon_{\ell_1^\pm \ell_1^\mp}^A \simeq \dfrac{1}{2} \, \varepsilon_{B-L} \left( \dfrac{ Q_\text{VLL}^{B-L} \Lambda_{\rm E}^2 - 1}{1+\Lambda_{\rm E}^2} \right) \, ,
\end{equation}
which are now suppressed if one considers $Q_\text{VLL}^{B-L} \Lambda_{\rm E}^2 \simeq 1$. Large $B-L$ charges now require small $\Lambda_{\rm E}$. Thus the axial coupling becomes very small, while for the vector one we have the relation $\varepsilon_{\ell^\pm_{1}\ell^\mp_{1}}^V \simeq -\varepsilon$.

One may now wonder whether the allowed ranges for $\varepsilon_{n,p}^V$, unaltered by the addition of VLLs, can be simultaneously satisfied with the ones for the first generation $\varepsilon^V_{ee}$ when $Q_\text{VLL}^{B-L} \Lambda_{\L,\rm E}^2 \simeq 1$. In Fig.~\ref{fig:limits_on_couplings}, we show the experimental limits of the $Z^\prime$ couplings to the neutron, proton and electron, in the $\left(\varepsilon,\varepsilon_{B-L}\right)$ plane, using both non-updated (left panel) and updated (right panel) values from Table~\ref{tab:couplings}. The horizontal regions directly reflect the $\varepsilon_n^V=\varepsilon_{B-L}$ ranges shown in the first line (third column) of that table. Instead, the $\varepsilon_{ee}$ intervals in the third line/column, together with the relation $\varepsilon_{ee}\simeq -\varepsilon$ imposed by $Q_\text{VLL}^{B-L} \Lambda_{\L, \rm E}^2 \simeq 1$, define the vertical blue-shaded bands. Finally, the green-shaded bands stem from $\varepsilon_p^V=\varepsilon+\varepsilon_{B-L}$, taking into account the ranges for $\varepsilon_p^V$ given in Table~\ref{tab:couplings}. While there is a small overlap of the three regions in the case where the non-updated intervals are used (left panel, marked in purple), this is not the case with the updated results (right panel). This indicates a clear tension among all requirements needed for the proposed solution (small mixing with large $B-L$ charges) to work.

Even so, if one now turns to the second generation, neutrino-coupling suppression requires, once more, $Q_\text{VLL}^{B-L} \Lambda_\L^2 \simeq 1$, where $\Lambda_\L$ is now the mixing parameter of the second generation. Since we consider generation-independent $B-L$ charges, the limit $Q_\text{VLL}^{B-L} > 24$ still holds in this case. Suppression of the muon's axial coupling requires $Q_\text{VLL}^{B-L} \Lambda_{\rm E}^2 \simeq 1$, which implies that $\Lambda_\L \simeq \Lambda_{\rm E} < \sqrt{1/24} \simeq 0.2$. From Eq.~(\ref{eq:kappal}), one gets the bound $\kappa_\mu \gtrsim 0.92$, which agrees at $1\sigma$ with the experimental constraint on the Higgs-muon coupling modifier $\kappa_\mu=1.12^{+0.21}_{-0.22}$~\cite{CMS:2022dwd}, unlike the large mixing case.

\begin{figure}
  \centering
  \includegraphics[height=3.55cm]{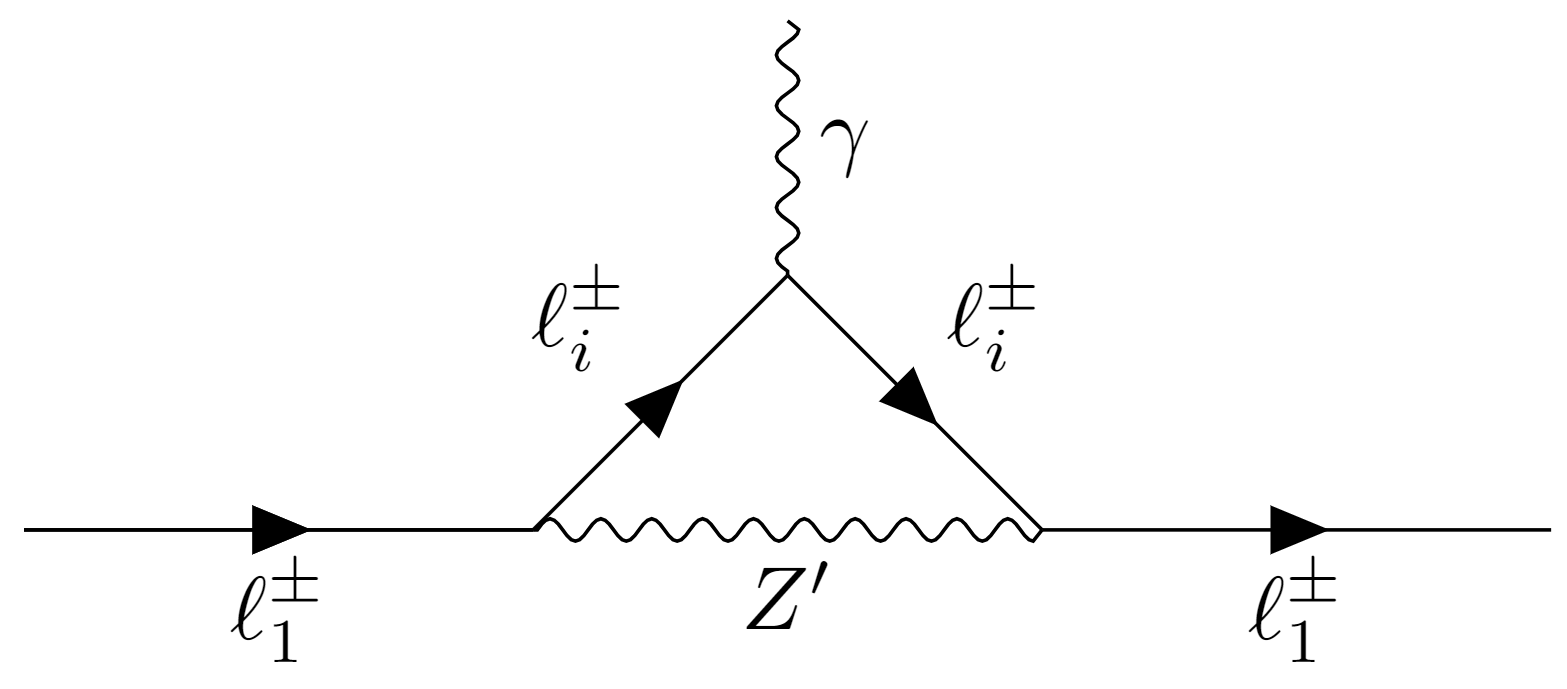}
  \hspace{0.1cm}
  \includegraphics[height=3.55cm]{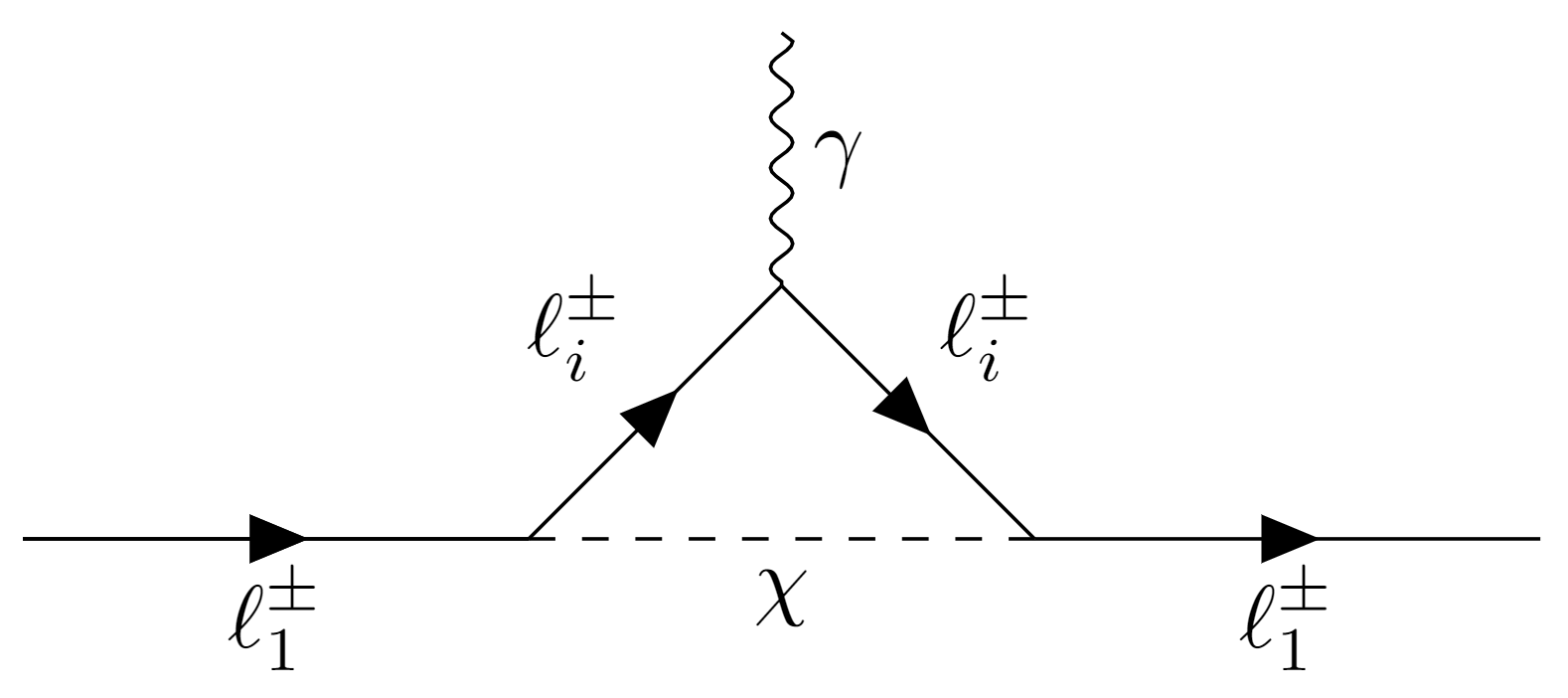}
  \caption{New one-loop contributions to the magnetic moment of $\ell_1^\pm$ (SM lepton of each generation) from diagrams involving the new heavy leptons together with the gauge boson $Z^\prime$ (left diagram) and the new scalar $\chi$ (right diagram). The fermionic index $i$ can represent the SM lepton, $i=1$, or the two new heavier fields, $i=2,3$.}
  \label{fig:feyndiagrams}
\end{figure}

Since we are now dealing with muons, it is imperative to compute the NP contributions to $(g-2)_\mu$ in the region that complies with both large $B-L$ charges and small mixing. The new one-loop diagrams are presented in Fig.~\ref{fig:feyndiagrams} and can be computed taking into account the general formulae given in Ref.~\cite{Dermisek:2021ajd}. In what follows, we work out the expressions for $\Delta a_\mu = \Delta a_\mu^{Z^\prime}+\Delta a_\mu^\chi$, under the previous assumptions for $Q_\text{VLL}^{B-L}$ and $\Lambda_{\L,\rm E}$. The first contribution ($\Delta a_\mu^{Z^\prime})$ coming from the diagram with the $Z^\prime$ reads
\begin{equation}
    \Delta a_\mu^{Z^\prime} \simeq \mathcal{A}_\mu + \mathcal{A}_\text{VLL} = -\dfrac{m_\mu^2 e^2}{8 \pi^2 m^2_{Z^\prime}} \left( \mathcal{C}_\mu \varepsilon^2 + \mathcal{C}_\text{VLL} \, \varepsilon^2_{B-L} Q_\text{VLL}^{B-L}  \right) \, , 
\end{equation}
where $\mathcal{A}_\mu$ ($\mathcal{A}_\text{VLL}$) corresponds to the case where the fermion in the loop is the muon (a new charged lepton). As for $\mathcal{C}_{\mu}$, we have 
\begin{equation}
    \mathcal{C}_\mu=2 \,  F_{Z^\prime} \left( \dfrac{m_\mu^2}{m_{Z^\prime}^2} \right) -  G_{Z^\prime} \left( \dfrac{m_\mu^2}{m_{Z^\prime}^2} \right) \simeq 0.321 \, ,
\end{equation}
with $F_{Z^\prime}$ and $G_{Z^\prime}$ being loop functions given by
\begin{align}
    F_{Z^\prime}(x) &= \dfrac{5x^4-14x^3+39x^2-38x+8-18x^2\log x}{12 (1-x)^4} \; ,\;G_{Z^\prime}(x) = - \dfrac{x^3+3x-4-6x\log x}{2 (1-x)^3} \, .
\end{align}
Given the small-mixing assumption between SM leptons and VLLs, and taking into account that there is no inter-generational VLL mixing, the masses of the new leptons are mostly given by the VLL bare mass terms $M_{\L,\rm E}$, as seen in Eq.~(\ref{eq:Mell}). Bearing in mind the conservative limits $M_\L \gtrsim 130~\text{GeV}$~\cite{Feng:2016ysn} and $M_{\rm E} > 200~\text{GeV}$~\cite{Dermisek:2021ajd}, and that $m_{Z^\prime} \ll M_{\L,\rm E}$, we can safely make the approximation
\begin{equation}
    \mathcal{C}_\text{VLL} = F_{Z^\prime} \left( \dfrac{M_\L^2}{m_{Z^\prime}^2} \right) + F_{Z^\prime} \left( \dfrac{M_{\rm E}^2}{m_{Z^\prime}^2} \right) \simeq 2 \, \lim_{x\to\infty} F_{Z^\prime} (x) = \dfrac{5}{6} \, .
\end{equation}
We point out that $\Delta a_\mu^{Z^\prime}$ is always negative, thus being in tension with experiment since it has the opposite sign of the current value for $\Delta a_\mu$.

The contribution stemming from the one-loop diagram with the new scalar in the loop is
\begin{equation}
    \Delta a_\mu^\chi \simeq \mathcal{A}_\L + \mathcal{A}_{\rm E} = \dfrac{m_\mu^2 e^2}{4 \pi^2 m^2_{Z^\prime}} \varepsilon^2_{B-L} \left( \mathcal{C}_\L \Lambda_\L^2 + \mathcal{C}_{\rm E} \Lambda_{\rm E}^2 \right) \quad , \quad
    \mathcal{C}_{\L,\rm E} = \dfrac{M_ {\L,\rm E}^2}{m^2_\chi} F_\chi \left( \dfrac{M_{\L,\rm E}^2}{m_\chi^2} \right)\,,
    \label{eq:deltamuchi}
\end{equation}
where
\begin{equation}
    F_{\chi}(x) = \dfrac{x^3-6x^2+3x+2+6x \log x}{6 (1-x)^4} \, .
\end{equation}
In order to evaluate the relevance of $\Delta a_\mu^\chi$, we recall the unitary limit on $m_{\chi}$ obtained in Eq.~(\ref{eq:limitMassChi}). Taking the lowest possible value for $\varepsilon_{B-L}$ from Table~\ref{tab:couplings}, the non-updated value $\varepsilon_{B-L}=2 \times 10^{-3}$ implies $m_\chi \lesssim 45~\text{GeV}$, as we have seen, while the updated one lowers the upper limit on $m_\chi$ to approximately $22~\text{GeV}$. Thus, one can consider the limit $ m^2_\chi / M_ {\L,\rm E}^2 \ll 1 $ and, to a good approximation, the coefficients $\mathcal{C}_{\L,\rm E}$ are
\begin{equation}
    \mathcal{C}_{\L,\rm E} \simeq \lim_{x\to\infty} x \, F_\chi(x) = \dfrac{1}{6} \, ,
\end{equation}
regardless of the value of $m_\chi$.

We now compute the total $\Delta a_\mu$ for a benchmark scenario. The $B-L$ charge is chosen to be the limit value $Q_\text{VLL}^{B-L} = 24$, from which we automatically get $\Lambda_\L=\Lambda_{\rm E}=\sqrt{1/24}$. We choose both $\varepsilon$ and $\varepsilon_{B-L}$ that satisfy the experimental bounds considering the left panel of Fig.~\ref{fig:limits_on_couplings}. Altogether, these amount to:
\begin{equation}
    Q_\text{VLL}^{B-L}=24 \quad , \quad \Lambda_\L = \Lambda_{\rm E} = \dfrac{1}{\sqrt{24}} \quad , \quad \varepsilon = - 1.5 \times 10^{-3} \quad , \quad \varepsilon_{B-L} = 2 \times 10^{-3} \, .
\end{equation}
These give rise to the following $\mathcal{A}$-contributions to $\Delta a_\mu$:
\begin{equation}
    \mathcal{A}_\mu \simeq -3.26 \times 10^{-8} \quad , \quad  \mathcal{A}_\text{VLL} \simeq -3.61 \times 10^{-6} \quad , \quad  \mathcal{A}_{\L, \rm E} \simeq 2.51 \times 10^{-9} \quad \Rightarrow \quad \Delta a_\mu \simeq -3.64 \times 10^{-6} \, .
\end{equation}
The $\mathcal{A}_\text{VLL}$ contribution dominates, however its value is three orders of magnitude deviated from $\Delta a_\mu$, and the sign is off, if one considers $\Delta a_\mu \sim 10^{-9}$.

On one hand, and regarding $\Delta a_\mu^{Z^\prime}$, $\mathcal{A}_\text{VLL}$ is expected to dominate for $Q_\text{VLL}^{B-L}>24$. Different values of $\varepsilon$ and $\varepsilon_{B-L}$ within the experimental bounds will not significantly alter the order of magnitude of each contribution. On the other hand, and looking at $\Delta a_\mu^\chi$, the maximum value for such contribution is the one already considered in the benchmark. Larger $B-L$ charges imply smaller $\Lambda_{\L, \rm E}$, which would yield smaller $\mathcal{A}_{\L,\rm E}$ contributions. Here, the same reasoning for different values of $\varepsilon_{B-L}$ applies. Despite the fact that the correct sign and order of magnitude can be obtained with $\Delta a_\mu^\chi$, such contribution is always subdominant in comparison with the $Z^\prime$ one, due to the term proportional to $Q_\text{VLL}^{B-L}$.

\begin{figure}
  \centering
  \includegraphics[height=8.5cm]{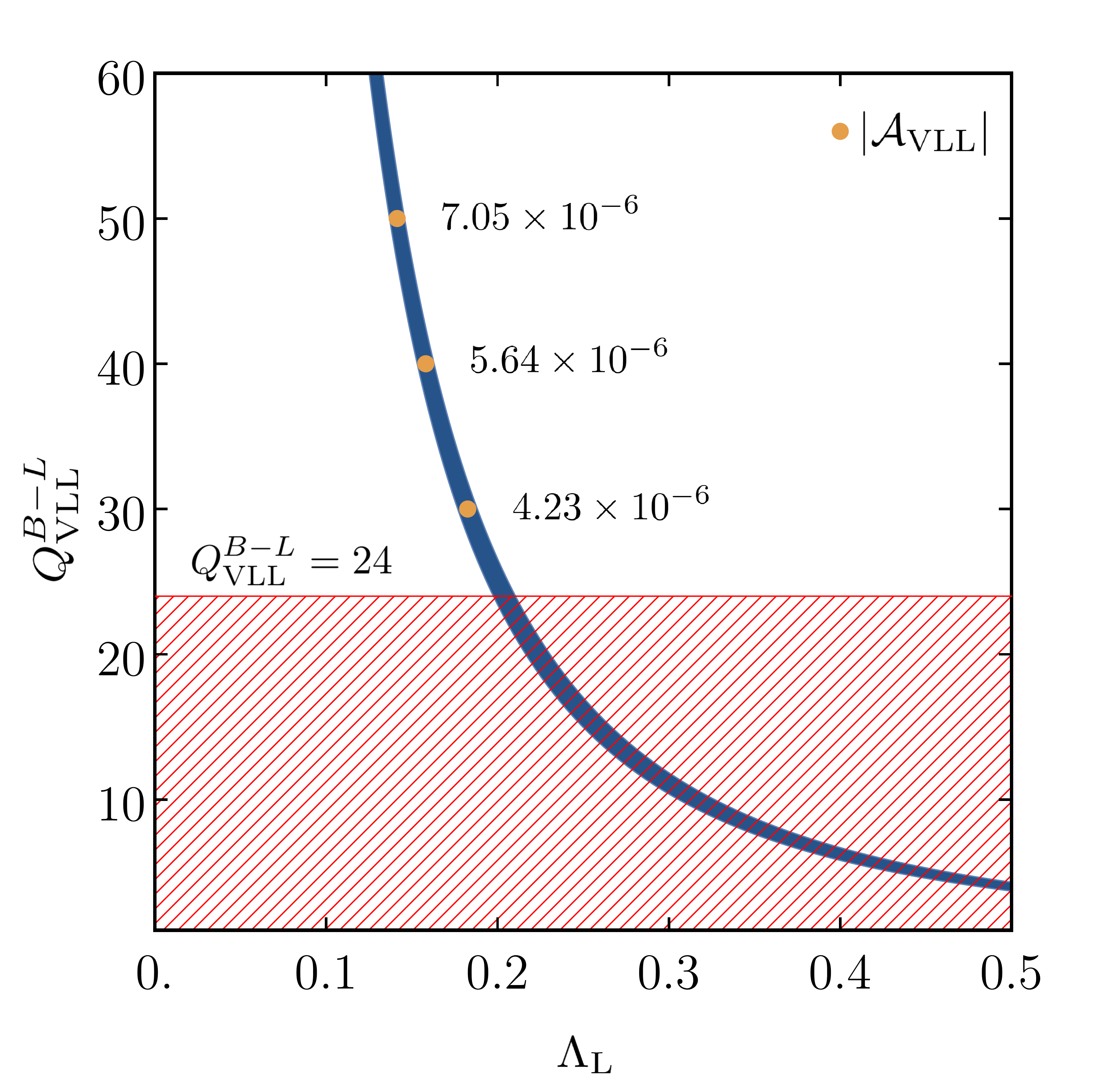}
  \caption{Neutrino coupling suppression in the plane $\left( \Lambda_\L,Q_\text{VLL}^{B-L} \right)$. The blue shaded region represents $|\varepsilon_{\nu_1 \nu_1}/\varepsilon_{B-L}|<10^{-3}$. In the allowed region in which $Q_\text{VLL}^{B-L}>24$ we show the absolute value of $\mathcal{A}_\text{VLL}$ (the dominant contribution to $\Delta a_\mu$) for $Q_\text{VLL}^{B-L}=30,40,50$, considering $\varepsilon_{B-L}=2 \times 10^{-3}$.}
  \label{fig:suppression_g-2}
\end{figure}

Our findings are summarised in Fig.~\ref{fig:suppression_g-2}. In the blue thin region of the $\left( \Lambda_\L , Q_\text{VLL}^{B-L} \right)$ plane, the neutrino couplings are suppressed, \textit{i.e.} $|\varepsilon_{\nu_1 \nu_1}/\varepsilon_{B-L}|<10^{-3}$. One can clearly see that, to achieve this for $Q_\text{VLL}^{B-L}=1$, the mixing has to be considerable -- large $\Lambda_\L$ is needed. If we decrease the mixing, then we need larger $B-L$ charges. We have seen that, from limiting the mixing in the first generation, the charge has to comply with $Q_\text{VLL}^{B-L}>24$, if one forgets the perturbativity limit of $\sqrt{4 \pi}$. This limit on the charge also applies to the second generation. From neutrino suppression, the upper bound $\Lambda_\L < \sqrt{1/24} \simeq 0.2$ follows. The region $Q_\text{VLL}^{B-L}<24$ is excluded in Fig.~\ref{fig:suppression_g-2}. It turns out that, in the allowed region, the dominant contribution to $\Delta a_\mu$ is, at least, three orders of magnitude higher than the value $\Delta a_\mu \sim 10^{-9}$, with a negative sign, opposite to what is required from experimental results. These values are shown for different $B-L$ charges by orange dots, assuming the minimum value for $\varepsilon_{B-L}$. If one had considered its maximum value, this would further increase. The choice $Q_\text{VLL}^{B-L}=24$ was a rather conservative one. We have seen that, in fact, assuming $\sqrt{4 \pi}$ as the perturbativity limit would increase the lower bound on the charge to $57$, which would lead to even larger values of $\Delta a_\mu$. Regarding the mass values, higher lower bounds for the heavy-lepton masses would lead to even larger values for $Q_\text{VLL}^{B-L}$, and consequently $\Delta a_\mu$.

In order to ascertain our findings, we performed a numerical scan of the model by now taking the $B-L$ charge of the vector-like fields as a new free parameter. We were able to numerically find $B-L$ charges as low as $63$, but most of the values obtained were much larger than this one, yielding values for $(g-2)_\mu$ several orders of magnitude off. We were not able to reproduce the current value for $\Delta a_\mu$ in our numerical scan.

\section{Concluding remarks}

In this work, we revisited the $U(1)_{B-L}$ solution to the ATOMKI nuclear anomalies, in which the $B-L$ gauge boson is the hypothetical $X(17)$ particle, proposed to explain such discrepancies. A first solution relied on adding VLLs with $Q^{B-L}_\text{VLL}=1$, in order to achieve neutrino-coupling suppression. However, such model requires large mixing between the SM and the new fields, and is therefore disfavoured. A second scenario was put forward in which the VLLs have large $B-L$ charges -- this allows to counterbalance small mixing and suppresses the couplings of $X(17)$ with neutrinos, as required. We have computed the contributions to the muon's $g-2$ in that case. The one-loop diagram with the $Z^\prime$ and the new leptons running in the loop dominates with a term that is proportional to the $B-L$ charge. We found that its sign is wrong (negative sign) as well as its absolute value is too large. Despite the several SM theoretical uncertainties that have plagued the $(g-2)_\mu$ anomaly, such large values for $\Delta a_\mu$ are unacceptable. 

We have also performed a numerical scan of the full model for both cases: large and small mixing. We concluded that, with $Q_\text{VLL}^{B-L}=1$, the large mixing needed for neutrino-coupling suppression is in conflict with experimental data, and we did not find any viable suppression in that case. Regarding the possibility of large $B-L$ charges, we confirmed that $\Delta a_\mu$ is larger than the experimental value by several orders of magnitude. Unnatural or fine-tuned solutions may still be possible but we did not find any of those in our analysis. Thus, to the best of our knowledge, a $B-L$ solution to the ATOMKI hint for NP appears to be increasingly farfetched and contrived. Recently, a combined solution to the ATOMKI, $(g-2)_\mu$ and MiniBooNE anomalies appeared in~\cite{Ghosh:2023dgk}. The authors considered $U(1)_H$ extensions of the Type-I 2HDM plus a scalar singlet. Despite the different gauge and scalar structures, the one-loop contributions to the $(g-2)_\mu$ are identical to the ones we have obtained, with new light scalar and vector particles in the loop. They confirm that both contributions have opposite signs, and only fine-tuning between them can reproduce the correct value for the muon’s magnetic moment in their scenario.

Several experimental efforts are being put forward in order to hopefully explore the still-allowed parameter space. In~\cite{Darme:2022zfw}, the experimental reach of the PADME experiment in looking for light bosons is discussed. Recently, the FASER Collaboration also released their first results~\cite{FASER:2023tle} and the relevant parameter space will be further tested in the future. The parameter space allowed to explain the anomaly through a vector boson has been constrained using leptonic decays of the charged pion, and there appears to be a tension with the pure-vector solution~\cite{Hostert:2023tkg}. Therefore, the axial-vector solution might be preferred. However, there are a lot of theoretical uncertainties regarding estimates for the axial nuclear matrix elements~\cite{Barducci:2022lqd}. Despite the fact that a clear picture is still missing, the ATOMKI Collaboration continues to present evidence for the existence of the $X(17)$ particle~\cite{Krasznahorkay:2023sax}. It is hoped that several experimental endeavours will eventually shed some light on this matter in the near future.

\section*{Acknowledgments}
We thank Jonathan Kriewald for carefully reading the manuscript. This work is supported by Funda\c{c}{\~a}o para a Ci{\^e}ncia e a Tecnologia (FCT, Portugal) through the projects UIDB/00777/2020, UIDP/00777/2020, UIDB/00618/2020, UIDP/00618/2020, CERN/FIS-PAR/0002/2021, CERN/FIS-PAR/0019/2021 and CERN/FIS-PAR/0025/2021. The work of B.L.G. is supported by the FCT PhD grant SFRH/BD/139165/2018. F.R.J. thanks the CERN Department of Theoretical Physics for hospitality and financial support during the preparation of this work.

\bibliography{Ref}

\end{document}